\documentclass[5p]{elsarticle}

\usepackage{amssymb}
\journal{TIPP09 Proceedings in NIMA}

\begin{document}

\begin{frontmatter}



\title{Highly segmented thin microstrip detector with data-driven fast readout}

%\author{}
%
%\address{}
%
%
%***************
%   Authors
%***************
%
\author[First]{Marco Bomben\corref{cor1}}
%\author[Second]{Tim B. Secondauthor}
\ead{Marco.Bomben@ts.infn.it}
\cortext[cor1]{Corresponding author, on behalf of the Slim5 collaboration.} 
%\author[First,Second]{James Q. Thirdauthor}
%\author[]{}
%\author[]{}
%
%
%***************
%   Addresses
%***************
%
\address[First]     {Universit\`a degli Studi di Trieste \& INFN-Sez. di Trieste \\
 Padriciano 99 - 34012 Trieste - Italy}
%\address[Second]     {Second affiliation, Address, City and Postcode, Country}
%\address[]{}
%\address[]{}

%
%===============================================================
%
%   Abstract
%
%===============================================================
%
%
\begin{abstract}
In September 2008 the Slim5~\cite{bib:slim5} collaboration submitted a
 low material budget
silicon demonstrator to test with 12 GeV/c protons, at the PS-T9
test-beam at CERN. Inside the reference telescope, two different detectors
were placed as device under test (DUT).
 The first was a 4k-Pixel Matrix of Deep N Well MAPS,
developed in a 130 nm CMOS Technology, providing digital sparsified
readout. The other one was a high resistivity double sided silicon detector,
with short strips at 45$^{\circ}$ angle to the detector's edge,
read out by the FSSR2 chip. The FSSR2 is a 128 channel data-driven fast
 readout chip developed by Fermilab and INFN.
In this paper we describe the main features of latter sensor, the striplet.
The primary goal of the test was to measure the efficiency and the
resolution of the striplets.
The data-driven approach of the FSSR2 readout chips has been fully exploited by  the DAQ system.
\end{abstract}

%
%
%===============================================================
%
%   Keywords
%
%===============================================================
%
%
\begin{keyword}
Striplets \sep striplet detector \sep charged particle tracking


\end{keyword}

\end{frontmatter}


%
%
%===============================================================
%
%   Main text

\section{Introduction}
Vertex detectors for experiments at future colliders such as
the SuperB Factory~\cite{bib:SuperB} or the International Linear
 Collider~\cite{bib:ILC}
will need to fulfill very stringent requirements on position resolution,
readout speed, material budget and radiation tolerance.
To address these request, a high resistivity double-sided strip detector has been designed and
 fabricated, 
featuring strips tilted by 45${^o}$ with respect to the detector edge, the 
 striplets. 
It is worth noting that in the SuperB 
Conceptual Design Report~\cite{SuperB:CDR} this 
 is the baseline solution proposed for the layer 0 of silicon vertex
 detector.
In this note we report about the sensor design, the readout chip we used, 
 the experimental setup of the test-beam, the data-analysis main features and 
 eventually the results for striplet performance, concerning space-point 
 resolution and hit efficiency.

\section{Striplet geometry}
\begin{figure}[h]
\centering
\includegraphics[width=0.45\textwidth]{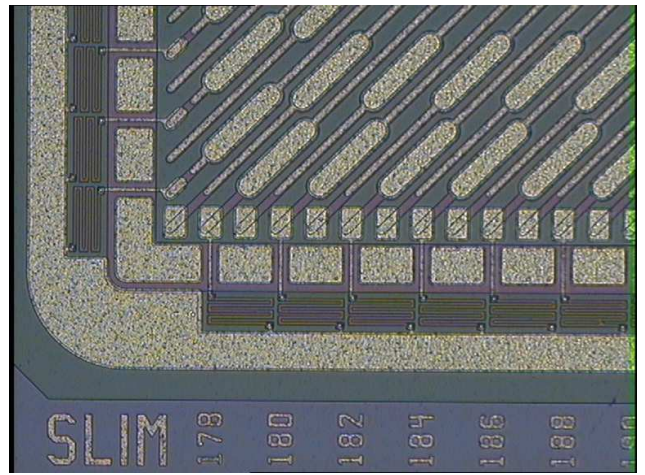}
\caption{Detail of a corner of the SLIM5 striplet detectors.}
\label{fig:striplets}
\end{figure}

The striplet geometry is optimized to limit the strip length ($18~$mm) and 
the material budget (double-sided, $200~\mu$m thin detector).
The strips are thus tilted by $45^\circ$ with respect to the detector edge, 
as shown in Fig.~\ref{fig:striplets}.
\label{STRIPLETS}
The short strip length reduces the average occupancy per channel 
and allows coping with the increased strip-to-backside capacitance.
The strips are AC coupled, with integrated capacitors and polisilicon 
biasing-resistors. In order to minimize the dead area along the edges, 
the resistors are placed outside the guard ring, a solution already 
adopted for the BaBar SVT sensors. 
The detector is fully depleted at $10~$V bias.

\section{The experimental set-up}
\label{EXPSETUP}
\begin{figure}[h]
\centering
\includegraphics[width=0.45\textwidth]{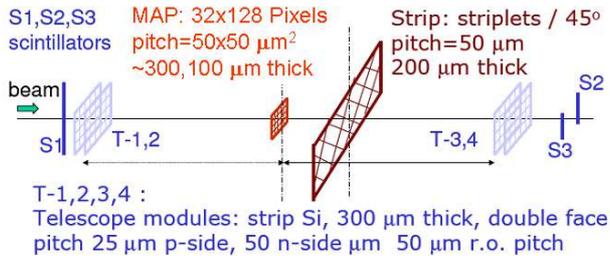}
\caption{The SLIM5 Beam Test Setup.
}
\label{fig:setup_testbeam}
\end{figure}

The Slim5 experiment was given the possibility of using for a beam-test 
the T9 facility at the CERN PS. 
To minimize the effect of multiple scattering on the resolution, 
we choose the maximum momentum that could be obtained, 
that is we got protons of 12 GeV/c as impinging particles, 
with spills of 400~ms and typical $10^4 \to 10^6$ particles/spill. 
The beam profile had a width of about 0.5~cm RMS. 
For the reference telescope, four silicon strip detectors, 
2cm$\times$2cm double-sided, AC coupled with 50 $\mu$m read-out pitch, have 
been used, one pair upstream and one downstream the device under test (DUT). 
The modules (telescope and DUTs) were placed on a customized motorized table
with remote control. 
The two pairs of telescope detectors were placed 40~cm apart, 
with a fixed 3.5~cm distance between the sensors inside each pair. 
The whole setup of the beam test is schematically shown in 
Fig.~\ref{fig:setup_testbeam}, where one can see also the scintillators, 
used for trigger studies.

\section{Striplets readout  and calibration}
%FSSR2 matches the striplet's cap. Calibration's results.

The striplet sensor is read out by the data-driven FSSR2 chip~\cite{bib:FSSR2}, 
the second release of the Fermilab Silicon Strip Readout Chip, 
derived from the one originally designed for the silicon pixel 
detector of the BTeV experiment.
The chip is completely data-driven and operated at 20~MHz readout clock, 
allowing a maximum data transmission rate of 240~Mbit/s over 
the six available transmission lines.
Each chip serves 128 strips, providing the address, the time stamp and a 
3 bit amplitude information for all hits.
A total of 384 channels were read out on each side of the sensor
by three FSSR2 chips, 
here used for the first time to process also negative signals.
Among the several programmable options we have chosen to operate the
chips at low gain, base line restorer selected, 125~ns shaping time,
a threshold corresponding to 4400(6300)~e$^-$ for p(n)-side respectively.
The same read out is used for the beam telescope detectors. 

For calibration purposes, an internal square-wave pulser provides voltage 
steps on the integrated injection capacitance of $40~$fF.
The injection capacitor discharge feeds the amplification chain.
We evaluate the input noise charge by measuring the fraction of hits over
 threshold as a 
function of the input charge at a given discriminator threshold. 
The results of the calibrations are summarized in Tab.~\ref{tb:FSSR2Calib}.
% p-side noise 630~ENC (S/N$\approx$25), gain $\approx$96~mV/fC.
% n-side noise 1020~ENC (S/N$\approx$16), gain $\approx$67~mV/fC.

\begin{table}[hbt]
\begin{center}
\caption{Calibration results for the striplet detectors.
\label{tb:FSSR2Calib}}
\vspace{\baselineskip}
\begin{tabular}{|c|c|c|}
\hline

Polarity & positive  & negative \\

\hline
Noise (e$^-$)& 630       & 1020   \\
S/N      & 25        & 16         \\
Gain (mV/fC)& 96     & 67         \\
Thr.Dis. (e$^-$)& 880       & 780\\ \hline
\end{tabular}
\end{center}
\end{table}

\section{Data analysis}
\label{RESULTS}
% General Description of methods and tools
% few words but many figure to show results
\subsection{Track Reconstruction}
Data analysis starts with the reconstruction of tracks using the
telescope detectors.  Clusters of fired strips are formed using a
simple algorithm that simply associates adjacent fired strips.
Typically a true cluster (one coming from a passing particle) contains
one or two fired strips.  The position of the cluster is calculated by
weighting each fired strip with its measured deposited charge.  For
each telescope plane, clusters in measured on the U-coordinate
(horizontal) side of the detector are combined with the V-coordinate
(vertical) clusters to form space points.

The track parameters are the result of a  $\chi^2$-fit to the 
 telescope-determined space-points, under straight-line assumption.

\subsection{Detector alignment}

Precise alignment of the telescope detectors as well as the devices
under test is essential for making accurate determinations of detector
performance.  
The alignment algorithm is based on minimizing track-hit residuals as
a function of detector position.
The alignment procedure starts by reconstructing tracks using very
large roads and applying no $\chi^2$ cuts.
  The alignment parameters are extracted from linear fits to
these residuals.  The procedure is iterated until specified convergence
criteria are satisfied.  The mean of the hit-track residuals after
alignment is typically on the order of a few microns.

\subsection{Determination of efficiency and resolution}

Once the telescope detectors and the DUT(s) were aligned, we can
measure the efficiency and resolution of the DUTs.  We select events
with exactly one good reconstructed track and consider the
intersection point of the fitted track with one striplet detector as DUT,
 The DUT
will generally have a space point caused by the track itself (the
signal hit), but it can also have hits that arise from noise sources.
We distinguish signal hits from noise hits statistically by fitting
the distribution of the residual distribution, where by residual we
mean the position of the hit found in the DUT minus the position of
intersection of the extrapolated track.

The space-point resolution of the DUT is determined from the width of the
fitted residual distribution.  The contribution to this width of the
track extrapolation uncertainty and multiple scattering effects, both
typically around 5 $\mu m$, are subtracted (in quadrature) to yield
the intrinsic resolution.

\section{Striplets results}
%\subsubsection{Resolution S1,2}
%\subsubsection{Efficiency}

We analyzed runs with one striplet detector placed as DUT to determine 
 its hit-efficiency and space-point resolution.

For a track with normal incidence, a signal cluster 
contains one strip in 82.5(85.6)\% of the cases or two adiacent strips in 16.1(13.3)\%
the cases on U(V)-coordinate. 

 For each track that passes the fiducial area of the striplet DUT, we look 
 for a cluster close to the track-intercept in the DUT plane, 
 separately for p and n-side.
For the striplets more than 98\% of the tracks have a space point within 
80~$\mu$m from the intersection point of the track.

\begin{figure}[ht]
\centering
\includegraphics[width=.23\textwidth]{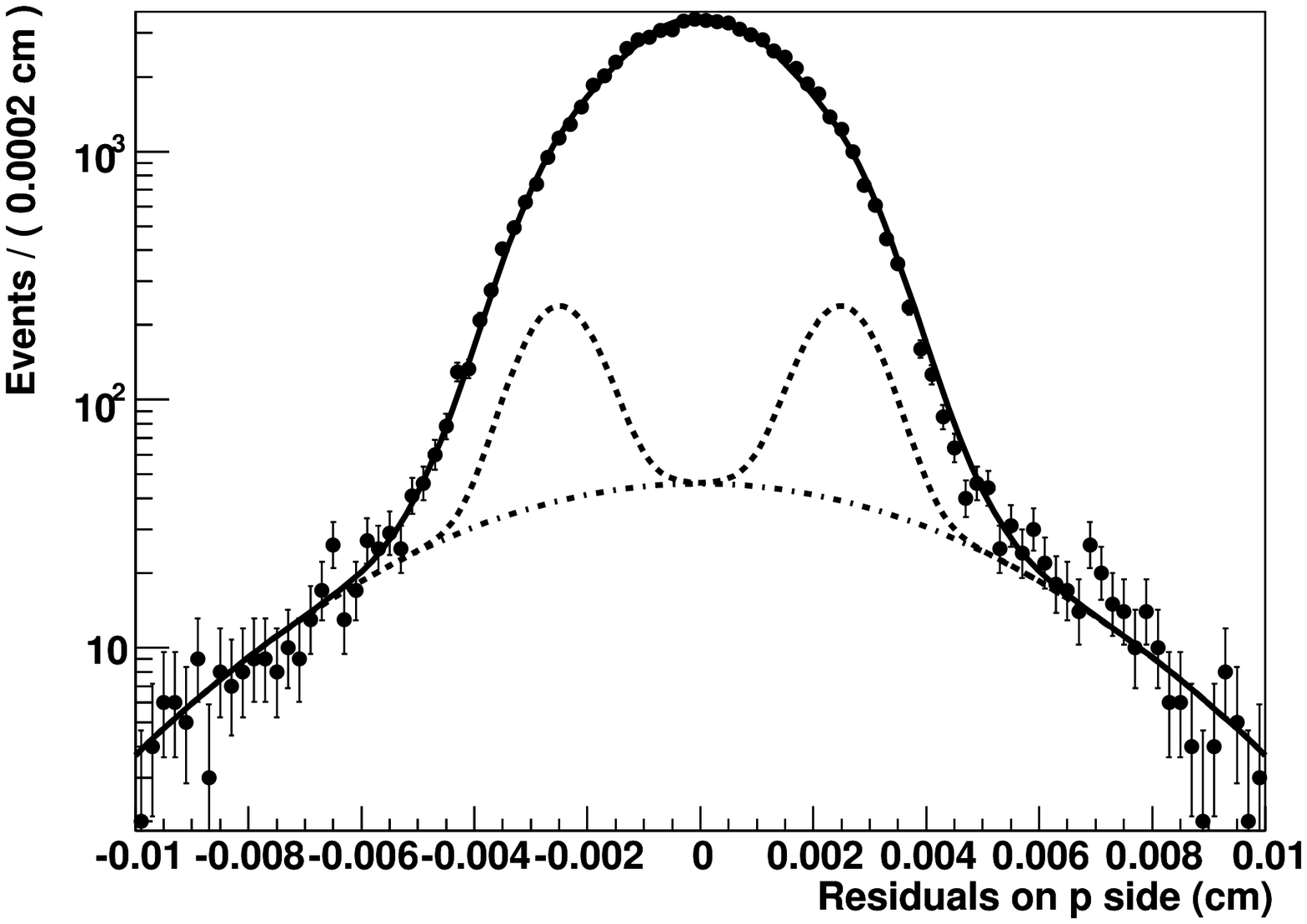}
\includegraphics[width=.23\textwidth]{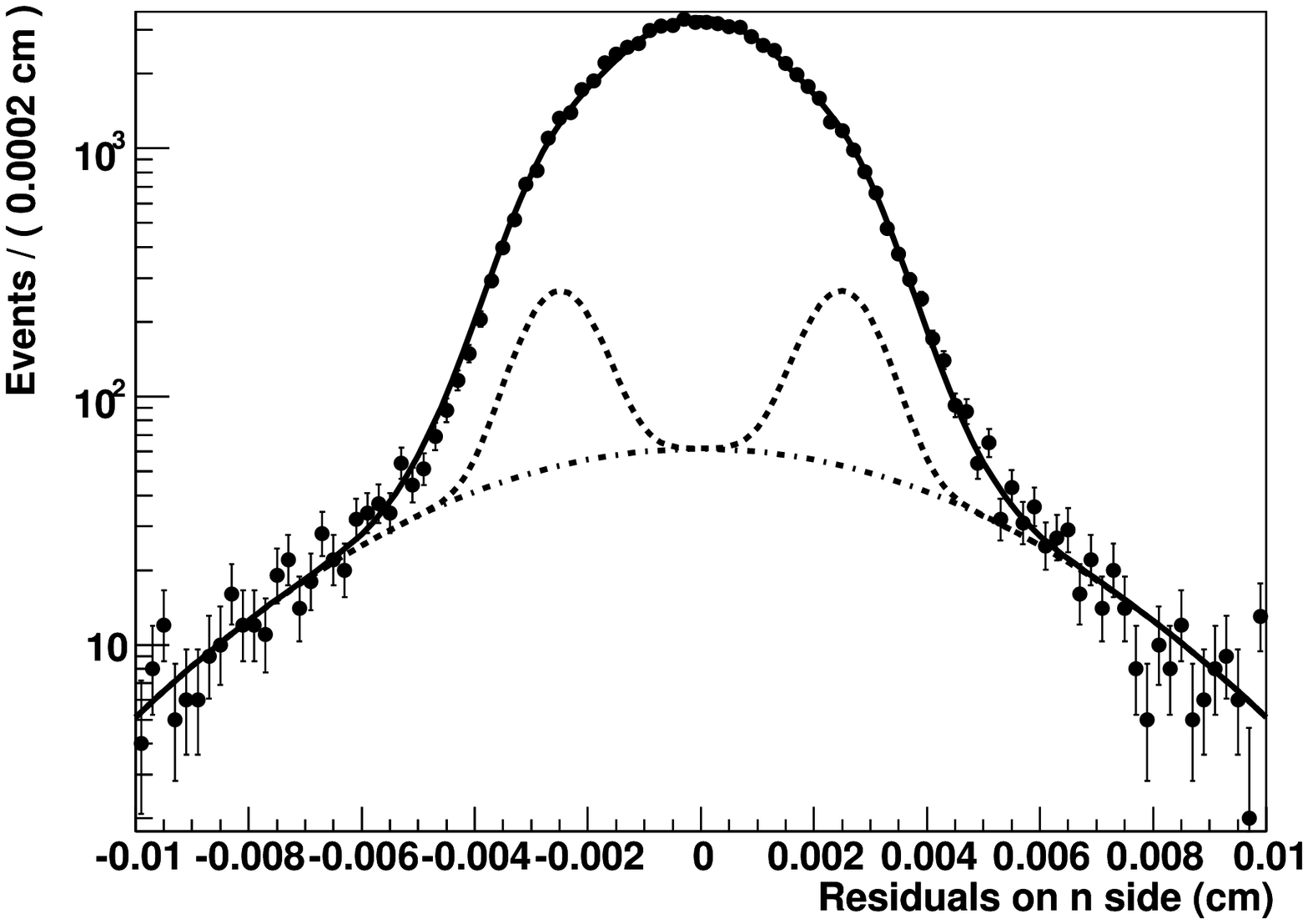}
\caption{Residual fit for the striplets as a DUT. The two plots correspond to
 fitting the p-side and the n-side, respectively.
 Dash-dotted line represent uncorrelated clusters; dashed represents true
 2-strip  clusters misreconstructed as 1-strip clusters.}
\label{fig:resid_fit_striplets}
\end{figure}

After alignment, the typical residual distribution are shown in
 figure~\ref{fig:resid_fit_striplets}. 
 We fit the residual distribution with a probability density function (p.d.f.)
 that is the sum of three functions: a) one central 
 gaussian, b) a double-peak function (modelled by two gaussians with same sigma 
 but opposite mean), and c) a wide gaussian, with mean fixed at zero.
Each function represents a specific class of events. 
 The central gaussian (a) represents correctly reconstructed hits.
 This kind of events represent more than the 90\% of the total; we use the 
 width of residual distribution to evaluate the space-point resolution.
 The double-peak function (b) represents true 2-strip cluster hits, wrongly 
 reconstructed as 1-strip cluster; this happens because of the rather high 
 value for threshold: a proton hitting the sensor roughly in the middle of 
 two strips generates an amount of charge that is shared almost equally 
 by the two adjacent strips. Due to signal and noise fluctuations, one 
 strip could fire and the other not. This leads to an error in the hit position
 of the order of $\pm 25 \mu m$. We fix the mean of the double-peak function to 
 this value. We find from the fit that the 5\% of the total events belong 
 to this category.
 The wide gaussian (c) describes noise hits (and so uncorrelated in space), 
 and also some signal from strips close to dead channels. This kind of 
 event accounts for $\sim$3\% of the total.

As said, the intrinsic resolution is obtained 
from the $\sigma$ of the central gaussian (a) and after the subtraction 
of the track extrapolation uncertainty and multiple scattering effects, 
yields 13.6 (14.1)~$\mu$m for the p(n)-side, slightly better than 
the digital resolution for a 50~$\mu$m read out pitch; this is due to the 
 high fraction of clusters with 1-strip only.
 It is expected a slight worse performance for n-side due to non-ideal 
 response of FSSR2 chip for negative input signals.

As a comparison, we estimated the full-width-half-maximum (fwhm)
 of the fitted p.d.f. 
 and its variance too. We tried also to fit the residual with 
 a single gaussian function.
 The results are reported in table~\ref{tab:rescomparison}.

\begin{table}[hbt]
\begin{center}
\caption{\label{tab:rescomparison}Comparison for central gaussian ($\sigma_C$), fwhm/2.35 and
 square root of variance ($\sqrt{\rm{V}}$)
 of p.d.f. used, and single-gauss width ($\sigma_S$) of residual distribution.}
\vspace{\baselineskip}
\begin{tabular}{|c| c| c| c| c|}
\hline
side & $\sigma_C$ ($\mu$m) & fwhm/2.35 ($\mu$m) & $\sqrt{\rm{V}}$ ($\mu$m)& $\sigma_S$ ($\mu$m)\\
\hline
p &  15.4 & 16.6 & 18.7 & 16.6 \\
n &  15.9 & 17.3 & 19.6 & 17.2 \\
\hline
\end{tabular}
\end{center}
\end{table}

It is clear from the comparison that the width of a single gaussian is wider 
 than the central one of the p.d.f., due to the presence of misreconstructed 
 hits; moreover, the fwhm (rescaled by 2.35) matches the
 single-gaussian width very well.

\section{Conclusions}
The Slim5 collaboration realized a double-sided microstrip silicon detector,
 with reduced thickness (200$\mu m$) and strips tilted by $\pm 45^\circ$ with 
 respect to the edge. The sensor was readout by a data-driven fast chip, the 
 FSSR2, used here for the first time to read negative signals.
 The whole detector was one of the DUTs in a test-beam at CERN
 PS-facility, in which several characteristics were measured.
 We find that the detector hit-efficiency was more than 98\%, with a 
 space-point resolution ($\sim 14 \mu$m) better than the digital 
  resolution for a 50~$\mu$m read out pitch.

%
%===============================================================
%
%   References
%
%===============================================================
%
%


\begin{thebibliography}{00}

%\bibitem{}
%

\bibitem{bib:slim5}
SLIM5 Collaboration - Silicon detectors with Low Interaction with Material, 
{\tt http://www.pi.infn.it/slim5/}

\bibitem{bib:SuperB}
For an introduction: {\tt http://www.pi.infn.it/SuperB}

\bibitem{bib:ILC}
For an introduction: {\tt http://www.linearcollider.org}

\bibitem{SuperB:CDR}
The SuperB Concptual Design Report, INFN/AE-07/02, SLAC-R-856, LAL 07-15, Available online at:
{\tt http://www.pi.infn.it/SuperB}

\bibitem{bib:FSSR2}
V.~Re {\it et al.}
IEEE Transactions on Nuclear Science, vol. 53, issue 4, pp. 2470-2476


%
\end{thebibliography}
\end{document}